\documentclass[11pt,superscriptaddress,aps,prd,preprint]{revtex4}
\usepackage[active]{srcltx}
\usepackage{graphicx}
\usepackage[utf8]{inputenc}
\usepackage{amsmath}
\usepackage{xcolor}
\usepackage{subcaption}
\usepackage{times,txfonts}
\usepackage{float}%para fixar as figuras na posição desejada

\begin{document}

\title{Dirac equation on a catenoid bridge: a supersymmetric approach}

\author{\"{O}. Ye\c{s}ilta\c{s}}
\affiliation{Department of Physics, Faculty of Science, Gazi University, 06500 Ankara, Turkey}
\email{yesiltas@gazi.edu.tr}

\author{J. Furtado}
\affiliation{Universidade Federal do Cariri(UFCA), Av. Tenente Raimundo Rocha, \\ Cidade Universit\'{a}ria, Juazeiro do Norte, Cear\'{a}, CEP 63048-080, Brasil}
%\email{job.furtado@ufca.edu.br}

\author{J.E.G. Silva}
\affiliation{Universidade Federal do Cariri(UFCA), Av. Tenente Raimundo Rocha, \\ Cidade Universit\'{a}ria, Juazeiro do Norte, Cear\'{a}, CEP 63048-080, Brasil}
%\email{euclides.silva@ufca.edu.br}

\begin{abstract}
In this paper we study the Dirac equation for an electron constrained to move on a catenoid surface. We decoupled the two components of the spinor and obtained two Klein-Gordon-like equations. Analytical solutions were obtained using supersymetric quantum mechanics for two cases, namely, the constant Fermi velocity  and the position dependent Fermi velocity cases.
\end{abstract}

\maketitle

\section{Introduction}

The geometry of the graphene layer \cite{geim,novoselov,katsnelson} plays an important role in the electronic structure. The possibility of developing new electronic devices due to the effect of graphene's curvature is promoting a growing interest in quantum mechanics at curved surfaces. Electronic properties of a graphene nanotorus were studied considering the dynamics governed by the Schr\"{o}dinger equation \cite{GomesSilva:2020fxo}, \cite{dainamle} as well as by Dirac equation \cite{Yesiltas:2018zoy}, \cite{dainam2}. Geometry fluctuations produce  pseudomagnetic fields \cite{ribbons} whose effects can be seen at ripples \cite{contijo} and corrugated layers \cite{corrugated}. The curvature at the tip of a conical layer produces a topological phase \cite{furtado}, whereas helical strips induce chiral properties \cite{dandoloff1,atanasovhelicoid,atanasov}. 

Another interesting geometry studied in the last years is the catenoid geometry. As a minimal surface,  two dimensional wormhole geometry is equivalent to a catenoid which are possible solutions for traversable wormholes \cite{Wo}. In Ref. \cite{wormhole,picak} a bridge connecting a bilayer graphene was proposed using a nanotube. In order to obtain a smooth bridge,  Ref.\cite{Dandoloff}, \cite{dandoloff} suggested a catenoid surface to describe the bilayer and the bridge using only one surface. This can be achieved due to the catenoid curvature which is concentrated around the bridge and vanishes asymptotically \cite{spivak}. For a non-relativistic electron, the surface curvature induces a geometric potential in the Schr\"{o}dinger equation. The effects of the geometry and external electric and magnetic fields upon the graphene catenoid bridge was explored in Ref.\cite{euclides}, where a  single electron is governed by the Schr\"{o}dinger equation on the surface. Incidentally, the influence of a position-dependent mass problem upon the electron on a catenoid bridge was studied in Ref. \cite{euclides2}, where it was proposed an isotropic position-dependent mass as a function of the Gaussian and mean curvatures.

The supersymmetry (SUSY) arose as an effort to obtain an unified description of all interactions of nature \cite{wess}, but it was immediately realized that it provides powerful techniques to obtain analytical solutions for problems in several branches of physics \cite{s1}. The factorization method was first introduced by Dirac in order to obtain the spectrum of the quantum harmonic oscillator \cite{dirac}. Later, Schr\"{o}dinger applied the same method for the radial part of the Coulomb problem \cite{schrodinger}. The supersymmetric quantum mechanics (SUSY QM) is an intense field of research and nowadays there is a wide set of Hamiltonians which are  analytically solvable through the factorization method \cite{sukhatme}. On the other hand, there are more representations of SUSY QM such as shape invariance \cite{R}, Darboux transformations \cite{F} and Hamilton  hierarchies \cite{DJ}. SUSY QM has  received lot of interest for   beautiful mathematical insight in both relativistic and non-relativistic quantum mechanics \cite{1,2,3}.

In this paper we study some properties of a relativistic electron constrained to move on a catenoid surface. Unlike the non-relativistic electron, the Dirac electron does not couple directly to the surface curvature. Instead, the momentum operator (covariant derivative) is modified by the surface tetrads and connection 1-forms.  In section II, we study the massless Dirac equation on the catenoid and we decoupled the two components of the spinor. As a result, we obtain two second-order Klein-Gordon-like equations with an effective potential depending on the surface curvature and on the effective Fermi velocity. In section III we present briefly the SUSY QM approach which was used to find analytical solutions for the Dirac equation on the catenoid in two particular cases, namely, the constant Fermi velocity case and the position-dependent Fermi velocity case. In section IV we present our final remarks.

\section{Massless Dirac equation on a catenoid}
\label{section1}

\begin{figure}[h!]
    \centering
    \includegraphics[scale=0.5]{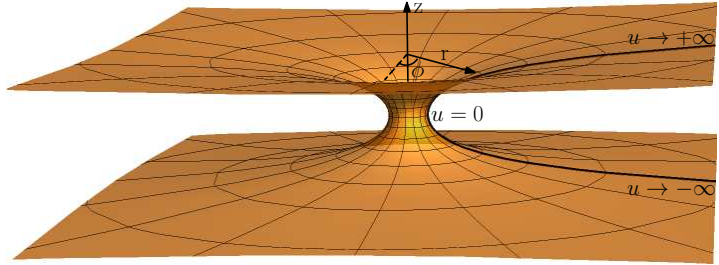}
    \caption{Catenoid}
    \label{Fig1}
\end{figure}

The catenoid can be described by the meridian $u\in (-\infty, \infty)$ and the parallel $\phi\in [0,2\pi)$, as
\begin{equation}
\label{coordinates}
    \vec{r}=\sqrt{R^2 +u^2}(\cos\phi\hat{i}+\sin\phi\hat{j})+R \sinh^{-1}\left(u/R\right)\hat{k},
\end{equation}
where $R$ is the radius of the catenoid bridge and $u=u(z)=R\sinh(z/R)$ , as depicted in Fig.(\ref{Fig1}). Using this coordinate system, the $2+1$ spacetime  interval reads
\begin{equation}
    \label{catenoidmetric}
    ds^2=dt^2 - du^2-(R^2+u^2)d\phi^2.
\end{equation}
%so that the induced metric on the catenoid is $g_{uu}=1$ and $g_{\phi\phi}=R^2+u^2$. In the vielbein formalism, by choosing the dreinbeins as $v^0 = dt$, $v^1=du$ and $v^2=\sqrt{R^2+u^2}d\phi$, the metric (\ref{catenoidmetric}) can be cast in the form $ds^2 = \eta_{ab}v^{a}\otimes v^{b}$, where $\eta_{ab}=diag(+1,-1,-1)$.

%From Eq.\ref{coordinates}, , where
Consider an orthogonal moving frame defined by the tangent vectors $\hat{e}_1 = \frac{\partial \vec{r}}{\partial u}$, $\hat{e}_2 =\frac{1}{\sqrt{R^2 + u^2}}\frac{\partial\vec{r}}{\partial\phi}$ and the normal unit vector $\hat{e}_3 =\hat{e}_1 \times \hat{e}_2$. The respective co-frame is given by $\omega^0 = dt$, $\omega^1 = du$, $\omega^2=\sqrt{R^2 + u^2}d\phi$ and $\omega^3 =0$, where $\omega^{a}=e^{a}_\mu dx^\mu$. Thus, $d\vec{r}=\omega^{1}\otimes\hat{e}_1 + \omega^{2}\otimes\hat{e}_2$ and the metric can be rewritten as $ds^2 = \eta_{ab}\omega^{a}\otimes \omega^{b}$, being $\eta_{ab}=diag(+1,-1,-1)$.
Using the torsion-free Cartan equation $T^a=d\omega^a+\omega^a_b\wedge \omega^b=0$ leads to the one-form connections $\omega^{a}_b = \Gamma_{cb}^{a}\omega^c$ of form $\omega^{1}_3 =\frac{R}{R^2 +u^2}\omega^1$,
$\omega^{2}_3 =-\frac{R}{R^2 +u^2}\omega^2$ and $\omega^{2}_1 =\frac{u}{R^2 +u^2}\omega^2$. Since 
$d\hat{e}_3 = \omega^{1}_3 \vec{e}_1 + \omega^{2}_3 \vec{e}_2=K_{b}^{a}\omega^{a}\otimes \vec{e}_b$, then the second fundamental form $K^{a}_b$ has components
\begin{eqnarray}
K^{a}_b &=&\frac{R}{R^2 +u^2}\left(\begin{array}{cc}
1 & 0 \\
0 & -1
\end{array}\right).
\end{eqnarray}
Therefore, the catenoid has gaussian and mean curvatures
\begin{eqnarray}
K=det(K^{a}_b)=-\frac{R^2}{(R^2 +u^2)^2}    &,& M=\frac{1}{2}K^{a}_a =0.
\end{eqnarray}
Accordingly, the catenoid has negative gaussian curvature vanishing far from the the bridge throat.
%\begin{eqnarray}
%T^1&=&dv^1+\omega^1_2\wedge v^2\\
%T^2&=&dv^2+\omega^2_1\wedge v^1,
%\end{eqnarray}

Intrinsically, only the $\omega^{2}_1$ is non zero. Using the metric compactibility condition $\omega_{ab} + \omega_{ba}=0$ and using $\omega^{12}=\omega^{12}_{\mu}dx^{\mu}$, then
\begin{eqnarray}\label{G12f}
\omega^{12}_{\phi}=\frac{u}{\sqrt{R^2+u^2}}.
\end{eqnarray}

The massless Dirac equation on a surface reads
\begin{equation}
    i v_F \hbar D_\mu \Psi =0,
\end{equation}
where $v_F$ is the Fermi velocity. The covariant spinor derivative is $D_\mu = \partial_\mu - \Gamma_\mu$, where $\Gamma_{\mu}=\frac{1}{4}\Gamma^{ab}_{\mu}\gamma_a\gamma_b$ is the spinor connection.
From Eq.(\ref{G12f}) the only non-vanishing spinor connection component is $\Gamma_{\phi}=\frac{1}{4}\Gamma^{12}_{\phi}\gamma_1\gamma_2$.
In (2+1)-D we can adopt the following representation for the flat $\gamma^{a}$ matrices 
$\gamma_0=\sigma_3$, $\gamma_1=-i\sigma_2$ and $\gamma_2=-i\sigma_1$.
The curved $\gamma^{\mu}$ matrices are related to the flat $\gamma^{a}$ matrices by $\gamma^{\mu}=e^{\mu}_a\gamma^a$, so that, 
\begin{eqnarray}
\gamma^t&=&e^t_0\gamma^0=\gamma_0\\
\gamma^u&=&\gamma^1\\
\gamma^\phi&=&e_2^{\phi}\gamma^2=\frac{1}{\sqrt{R^2+u^2}}\gamma^2,
\end{eqnarray}
and the spinor connection is
\begin{equation}
    \Gamma_{\phi}=\frac{i}{2}\frac{u}{\sqrt{R^2+u^2}}\gamma_0.
\end{equation}
The massless Dirac equation takes the form
\begin{eqnarray}\label{Diracequation}
\nonumber i\hbar v_F \left(\gamma^0\partial_0+\gamma^1\partial_u+\frac{1}{\sqrt{R^2+u^2}}\gamma^2\partial_{\phi}-i\frac{u}{2(R^2+u^2)}\gamma^2 \sigma_3 \right)\Psi=0.\\
\end{eqnarray}
Therefore, the catenoid geometry leads to a geometric potential as $V_1(u)=\frac{u}{4(R^2+u^2)}\gamma_1$,
which vanishes asymptotically and exhibits a parity odd behaviour near the throat.
%, as shown in fig.\ref{Diracoscilator}
%When $u\rightarrow\infty$, then $V_1(u)=\frac{1}{4u}$ and when $u\rightarrow 0$ %or $u<<R$ we have 
%\begin{eqnarray}
%V_1(u)=\frac{u}{4R^2}\left(1-\frac{u^2}{R^2}+\cdots\right).
%\end{eqnarray}
%Notice that the behaviour of the potential (\ref{pot1}) around the origin of the coordinate system resembles a Dirac oscilator, as we can see in fig. (\ref{Diracoscilator}).
%\begin{figure}[ht!]
 %   \centering
  %  \includegraphics[scale=0.9]{Pot_dirac.eps}
   % \caption{Potential (\ref{pot1}) for $R=10$. }
    %\label{Diracoscilator}
%\end{figure}
%Besides, the potential $V_1(u)$ is extremized when $u=\pm R$. 
The Dirac equation can be written as

\begin{eqnarray}
i\hbar \frac{\partial\Psi}{\partial t}= H_D \Psi,
\end{eqnarray}
where the Hamiltonian is given by
\begin{equation}
    H_D=-i\hbar v_F \left(\sigma_1 \left(\partial_u - \frac{u}{2(R^2 + u^2)}\right)-\sigma_2 \frac{1}{\sqrt{R^2 +u^2}}\partial_\phi\right).
\end{equation}
The independence of the catenoid metric with respect to time and the angular coordinate $\phi$ leads to the conservation of energy and of the angular momentum along the direction $z$. Thus, the wave function can be written as $\Psi(t,u,\phi)=e^{i\frac{E}{\hbar}t}e^{im\phi}\psi(u)$ what leads to the Dirac hamiltonian
\begin{eqnarray}
\label{hamiltonianu}
\nonumber H_D &=&-i\hbar v_F \left(\begin{array}{cc}
0 & \partial_u - \frac{u}{2(R^2 +u^2)}-\frac{m}{\sqrt{R^2 + u^2}} \\
\partial_u - \frac{u}{2(R^2 +u^2)}+\frac{m}{\sqrt{R^2 + u^2}} & 0
\end{array}\right).\\
\end{eqnarray}
Using the Hamiltonian in eq.(\ref{hamiltonianu}), the Dirac equation leads to coupled equations for the two components of the spinor $\psi=\left(\begin{array}{cc}
    \psi_1  \\
   \psi_2 
\end{array}\right)$, as
\begin{eqnarray}
\label{psi1}\left(\partial_u - \frac{u}{2(R^2 +u^2)}+\frac{m}{\sqrt{R^2+u^2}}\right)\psi_1 &=& -i\frac{E}{\hbar v_F}\psi_2\\
\label{psi2}\left(\partial_u - \frac{u}{2(R^2 +u^2)}-\frac{m}{\sqrt{R^2+u^2}}\right)\psi_2 &=& -i\frac{E}{\hbar v_F}\psi_1.
\label{coupleddiracequation}
\end{eqnarray}
Decoupling the system of first-order equations in eq. (\ref{coupleddiracequation}) leads to 
\begin{eqnarray}
-\psi_1''+\frac{u}{R^2 + u^2}\psi_1' +U_1 \psi_1 &=& \frac{E^2}{\hbar^2 v_{F}^2}\psi_1,
\end{eqnarray}
where $U_1 = \left(\frac{u}{R^2 + u^2}\right)' - \left(\frac{u}{R^2 + u^2}\right)^2 +\left(\frac{m}{\sqrt{R^2 + u^2}}\right)^2-\left(\frac{m}{\sqrt{R^2 + u^2}}\right)'$ and the prime stands for the derivative with respect to $u$. Employing the change on the wave function of the form $\psi_1 = (R^2 +u^2)^{1/4}\chi_1$ we obtain
\begin{eqnarray}
-\chi_1''+V_{eff,1}\chi_1 &=& \frac{E^2}{\hbar^2 v_{F}^2}\chi_1, 
\end{eqnarray}
where the effective potential $V_{eff,1}$ is given by

\begin{eqnarray}\label{pot1}
V_{eff,1}(u)=\Bigg[\left(\frac{m}{\sqrt{R^2 + u^2}}\right)^2-\left(\frac{m}{\sqrt{R^2 + u^2}}\right)'\Bigg]
%\frac{m^2}{R^2+u^2}+\frac{mu}{(R^2+u^2)^{3/2}}.
\end{eqnarray}
Analogously for the second component of the spinor we have:
\begin{eqnarray}
-\chi_2''(u)+V_{eff,2}(u)\chi_2=\frac{E^2}{\hbar^2 v_{F}^2}\chi_2(u),
\end{eqnarray}
where the potential $V_{eff,2}(u)$ is exactly like $V_{eff,1}(u)$ but under the interchanging $m\rightarrow -m$. The above equations were obtained under the consideration of a constant Fermi velocity $v_F$. The position dependent Fermi velocity case will be addressed in the following subsection.

%\begin{figure}[h!]
%    \centering
%    \includegraphics[scale=0.8]{p11.eps}
%    \caption{Effective potential for $R=10$ and four values of orbital angular momentum, namely, $m=\pm 1$ and $m=\pm 2$.}
%    \label{fig:my_label}
%\end{figure}

\subsection{Position-dependent Fermi velocity}

It is interesting to consider a position-dependent Fermi velocity, i.e., $v_F=v_F(u)$. The dependence of the Fermi velocity as a function only of $u$ lies in the symmetry of the catenoid on the angular variable, so that no dependence of $\phi$ is expected. Starting from (\ref{psi1}) and (\ref{psi2}), the process of decoupling the two equations renders for $\psi_1$
\begin{eqnarray}
\nonumber\psi_1''+\left[\Sigma(u)-\left(\frac{1}{v_F(u)}\right)'v_F(u)+\Lambda(u)\right]\psi_1'+\left[\Sigma(u)'-\left(\frac{1}{v_F(u)}\right)'v_F(u)\Sigma(u)+\Lambda(u)\Sigma(u)\right]\psi_1=-\frac{E^2}{\hbar^2 v_F(u)^2}\psi_1\\
\end{eqnarray}
where,

\begin{eqnarray}
\Sigma(u)&=&\frac{m}{\sqrt{R^2+u^2}}-\frac{u}{2(R^2+u^2)},\\
\Lambda(u)&=&-\frac{m}{\sqrt{R^2+u^2}}-\frac{u}{2(R^2+u^2)}.
\end{eqnarray}
In order to obtain a Klein-Gordon-like equation we must perform the following change of variables $\psi_1(u)=\kappa(u)\omega_1(u)$, imposing that $\omega_1(u)$ obeys a Sturm-Liouville equation. Such imposition leads to a condition on the function $\kappa(u)$ which states that 

\begin{equation}
    \kappa(u)=\exp\left[\frac{1}{4}\ln(R^2+u^2)-\frac{1}{2}\ln(v_F(u))\right].
\end{equation}
Hence, we obtain
\begin{eqnarray}\label{kg1}
-\omega_1(u)''+U_{eff,1}(u)\omega_1(u)=\frac{E^2}{\hbar^2 v_F(u)^2}\omega_1(u),
\end{eqnarray}
where,
\begin{eqnarray}
U_{eff,1}(u)=V_{eff,1}(u)+\bar{V}_{eff,1}(u),
\end{eqnarray}
being
\begin{equation}
    \bar{V}_{eff,1}(u)=-\frac{\left(v_F(u)'\right)^2 -2v_F(u) \left[\frac{2 m v_F(u)'}{\sqrt{R^2+u^2}}+v_F(u)''\right]}{4 v_F(u)^2}.
\end{equation}
Note that the potential $\bar{V}_{eff,1}(u)$ contains only the contributions from the position-dependent Fermi velocity, so that if we consider a constant Fermi velocity, $\bar{V}_{eff,1}(u)=0$ and we recover eq.(\ref{pot1}). Analogously, for the second component of the spinor we have an equation similar to eq.(\ref{kg1}) but with a potential $U_{eff,2}(u)$ equals to $U_{eff,1}(u)$ under the interchanging $m\rightarrow -m$.

\section{Supersymmetric Quantum Mechanics Approach}
Let us give a brief review of some aspects of supersymmetric quantum mechanics (SUSY QM) \cite{junker}
that we  use  in our search of exact solutions for the effective potential models given in the previous Sections. The techniques based on the factorization method which takes place in SUSY QM helps to identify the  Hamiltonians with solvable potentials. First, considering the one dimensional  eigenvalue equation below
\begin{equation}\label{susy1}
  \emph{H}~ \psi(z)=\left(-\frac{d^{2}}{dz^{2}}+U(z)\right)\psi(z)=E \psi(z)
\end{equation}
where $U(z)$ is a real scalar function. If we find two operators which are adjoint of each other and first order differental operators as
\begin{equation}\label{susy2}
  \mathcal{A}=\frac{d}{dz}+W(z),~~~~\mathcal{A^{\dag}}=-\frac{d}{dz}+W(z),
\end{equation}
where the superpotential $W(z)$ is a real function and element of the first order operators. Using the product of these operators, we can introduce two partner Hamiltonians defined by
\begin{eqnarray}\label{susy3}
 \emph{H}= \emph{H}_{1}&=&\mathcal{A}^{\dag}\mathcal{A}=-\frac{d^{2}}{dz^{2}}+U_{1}(z)\\
   \emph{H}_{2}&=& \mathcal{A} \mathcal{A}^{\dag}=-\frac{d^{2}}{dz^{2}}+U_{2}(z).
\end{eqnarray}
We can call the eigenfunctions of the Hamiltonians $\emph{H}_{1}$, $\emph{H}_{2}$ as $\psi^{(1)}_{n}$ and $\psi^{(2)}_{n}$ while eigenvalues as  $E^{(1)}_{n}$ and $E^{(2)}_{n}$ respectively. Then, one can introduce the partner potentials in terms of superpotentials as
\begin{equation}\label{susy4}
  U_{1}(z)=W(z)^{2}-W'(z),~~~~U_{2}(z)=W(z)^{2}+W'(z).
\end{equation}
These partner Hamiltonians are sharing the same energy eigenvalues except the groundstate.  In case of  unbroken SUSY, the ground state is not degenerate with zero energy $E^{(1)}_{0}$ which also leads to
\begin{equation}\label{susy5}
  \mathcal{A}\psi^{(1)}_{0}(z)=0.
\end{equation}
One can express the groundstate wavefunction in terms of the superpotential using (\ref{susy5})
\begin{equation}\label{susy6}
  \psi^{(n)}_{0}(z)=C_{n}\exp\left(\int^{z} W(t)dt\right).
\end{equation}
When it comes to the discrete spectrum of $H_{2}$, the relationship between the energies of these Hamiltonians are given by
\begin{equation}\label{susy7}
  E^{(2)}_{n}=E^{(1)}_{n+1},~~~~E^{(1)}_{0}=0.
\end{equation}
The eigenfunctions of the partner Hamiltonians are linked as below
\begin{equation}\label{susy8}
  \psi^{(2)}_{n}=\frac{1}{\sqrt{E^{(1)}_{n+1}}}\mathcal{A}\psi^{(1)}_{n+1},
\end{equation}
\begin{equation}\label{susy9}
  \psi^{(1)}_{n+1}=\frac{1}{\sqrt{E^{(2)}_{n}}}\mathcal{A}^{\dag}\psi^{(2)}_{n}.
\end{equation}
We remind that if $\psi^{(1)}_{n+1}$ is normalized, then, $\psi^{(2)}_{n}$ is also normalized(non-relativistic case). The annihilation of the groundstate of the operator $H_1$ means that there will be no SUSY partner for this state only \cite{sukhatme}.

\subsection{Constant Fermi velocity case }
Let us consider the effective potential model for the constant Fermi velocity given below
\begin{eqnarray}
  \emph{H}_{1} &=& -\frac{d^{2}}{du^{2}}+V_{1,eff}(u) \\
  V_{1,eff}(u) &=& \frac{m^2}{R^2+u^2}+\frac{mu}{(R^2+u^2)^{3/2}},
\end{eqnarray}
and
\begin{eqnarray}\label{susy9}
\emph{H}_{1} \chi^{(1)}_{n}=\epsilon^{2}\chi^{(1)}_{n}(u),
\end{eqnarray}
where $\epsilon=\frac{E}{v_F}$. The corresponding superpotential of the effective potential of the Hamiltonian $\emph{H}_{1} $ is
\begin{equation}\label{susy10}
  W(u)=\frac{m}{\sqrt{R^{2}+u^{2}}}.
\end{equation}
Then, one can obtain the groundstate wavefunction as
\begin{equation}\label{susy10}
  \chi^{(1)}_{0}=2^{-m}(u+\sqrt{R^{2}+u^{2}})^{-m}.
\end{equation}
On the other hand, the partner Hamiltonian is given by
\begin{eqnarray}\label{susy010}
  \emph{H}_{2} &=& -\frac{d^{2}}{du^{2}}+V_{2,eff}(u) \\
  V_{2,eff}(u) &=& \frac{m^2}{R^2+u^2}-\frac{mu}{(R^2+u^2)^{3/2}},
\end{eqnarray}
where we can write the first order operators $\mathcal{A}, \mathcal{A}^{\dag}$ as
\begin{eqnarray}\label{susy11}
  \mathcal{A} &=& \frac{d}{du}+ \frac{m}{\sqrt{R^{2}+u^{2}}}\\
  \mathcal{A}^{\dag} &=& - \frac{d}{du}+ \frac{m}{\sqrt{R^{2}+u^{2}}}. \label{susy12}
\end{eqnarray}
Next, we use a point transformation which is $u=R \tan x$ so that the equations (\ref{susy11}) and (\ref{susy12}) become
\begin{eqnarray}\label{susy13}
  \mathcal{A} &=& \frac{1}{R}\cos^{2} x \frac{d}{dx}+\frac{m}{R}\cos x\\
  \mathcal{A}^{\dag} &=& - \frac{1}{R}\cos^{2} x \frac{d}{dx}+\frac{m}{R}\cos x+\frac{\sin 2x}{R}.  \label{susy14}
\end{eqnarray}
We can obtain the transformed Hamiltonian $\emph{H}_{1}$ as
\begin{equation}\label{susy14}
  \emph{H}_{1}=-\frac{\cos^{4}x}{R^{2}}\frac{d^{2}}{dx^{2}}+\frac{2\cos^{2}x \sin2x}{R^{2}}\frac{d}{dx}+\frac{m\cos^{2}x(m+3\sin x)}{R^{2}}.
\end{equation}
Using $\chi^{(1)}_{n}(x)=\sec^{2}x \bar{\chi}^{(1)}_{n}(x)$ in (\ref{susy14}), we obtain
\begin{equation}\label{susy15}
  -\bar{\chi}''_{1}(x)+\xi\bar{\chi}_{1}(x)=0,
\end{equation}
with
\begin{equation}
    \xi=3m \sec x\tan x+(m^{2}+2)\sec^{2}x-4-\frac{E^{2}R^{2}}{v^{2}_{F}}\sec^{4}x,
\end{equation}
where we use $\chi^{(1)}_{n}(x)\rightarrow \chi_{1}(x)$ for the sake of simplicity. Let us analyze now two important cases, namely, the zero energy case and the case around the throat of the catenoid. \\
\subsubsection{Case I}
In case of zero energy, i.e. $\epsilon=\frac{ER}{v_F}=0$, we are lead to
\begin{equation}\label{susy16}
  -\bar{\chi}''_{1}(x)+(3m \sec x\tan x+(m^{2}+2)\sec^{2}x-4)\bar{\chi}_{1}(x)=0,
\end{equation}
where we can relate it to the system known as trigonometric Scarf-I potential in the literature \cite{sukhatme} whose eigenvalues and eigenfunctions are well known. Thus, one can get the solutions for zero energy as:
\begin{eqnarray}\label{susy181}
    \chi^{(1)}_{n}(x)&=&N_{1}~ [1+\exp(2ix)]^{2} \exp\left\{-2i[x-m \arctan \exp(ix)]\right\}.
\end{eqnarray}
 Let us  turn back to (\ref{susy15}) and apply $r=\sin x$ and $\bar{\chi}_{1}(r)=(r^{2}-1)^{1/4}\bar{\bar{\chi}}_{1}(r)$ to get
\begin{eqnarray}\label{susy22}
 -(1-r^{2}) \bar{\bar{\chi}}_{1}''(r)+2r\bar{\bar{\chi}}'_{1}(r)+\left(\frac{r^{2}-2}{4(1-r^{2})}+3m\frac{r}{(1-r^{2})}+
  \frac{m^{2}+2}{(1-r^{2})}-4-\frac{\epsilon^{2}}{(1-r^{2})^{2}}\right)\bar{\bar{\chi}}_{1}(r)=0.
\end{eqnarray}
 Because (\ref{susy22}) is not the type of hypergeometric  differential equation, we shall look at the approximation methods.\\ 
\textbf{Singularity analysis:}\\
Let us consider (\ref{susy22}) expressed below:
\begin{equation}\label{x}
   \bar{\chi}''_{1}(r)-\frac{2r}{1-r^{2}} \bar{\chi}'(r)-\frac{1}{(1-r^{2})^{2}}\left(\frac{r^{2}-2}{4}+3mr+(
  m^{2}+2)-4(1-r^2)-\frac{\epsilon^2}{1-r^2}\right)\bar{\chi}=0
\end{equation}
where $p(r)=-\frac{2r}{1-r^{2}}$ and $q(r)=-\frac{1}{(1-r^{2})^{2}} \left(\frac{r^{2}-2}{4}+3mr+
  m^{2}+2-4(1-r^2)-\frac{\epsilon^2}{1-r^2}\right)$. The singular points of (\ref{x}) are $r_0=\pm 1$. If $(r-r_0)p(r)$ and $(r-r_0)^2 q(r)$ are both analytic at $r=r_0$, then, $r_0$ is a 'regular singular point', if not, it becomes 'irregular singular point'. Hence, $r_0=1$ and $r_0=-1$ are irregular singular points because $q(r)$ is analytic at $r_0=\pm 1$, but $q(r)$ is not analytic due to the term $\frac{\epsilon^2}{1-r^2}$.  If $r_0$ is an irregular singular point, $(r-r_{0})^{k} p(x)$ and $(r-r_0)^{2k} q(r)$ are analytic where $k$ is the least integer number, then, the irregular singular point at $r=r_0$ has a rank of $k-1$. In our problem, $k=2$ and the rank is $1$. \\
  Next, we can continue to look for the solutions of (\ref{susy22}). %%%%%%%%%%%%%%%%%%%%%%%%%%%%%%%%%%%%%%%%%%%%%%%%%%%%%%%%%%%%%%%%%%%%%%%%%%%%%%%%%%%%%%%%%%%%%%%%%%%%%%%%%%%%%%%%%%%%%%%%%%%%%%%%%%%%%
\subsubsection{Case II: $r \rightarrow 0$ limit:}

Let us analyze the behavior of the solution of Eq.\ref{susy22} near the origin. By expanding the coefficients of Eq.\ref{susy22} in power-series up to first order in $r$, we obtain
\begin{equation}
    \chi''_{1} - 2r \chi'_{1}+ \left(\frac{10-4m^2 + 4\epsilon^2}{4}-3m r\right)\chi_1 =0,
\end{equation}
whose solution can be written as
\begin{equation}
    \chi_{1}(r) = c_1 e^{-\frac{3mr}{2}}H_{\alpha}\left(\frac{3m}{2}+r\right) + c_2 e^{-\frac{3mr}{2}} M\left(-\alpha/2,1/2,\left(\frac{3m}{2}+r\right)^2\right),
\end{equation}
where $\alpha=\frac{10+5m^2+4\epsilon^2}{8}$, $H_\alpha$ is the Hermite polynomial of degree $\alpha$ and $M\left(-\alpha/2,1/2,\left(\frac{3m}{2}+r\right)^2\right)$ is the Kummer confluent hypergeometric function. The polynomial degree $\alpha$ is a positive integer $\alpha=n$ provided that,
\begin{equation}
    \epsilon = \frac{\sqrt{8n-5(2+m^2)}}{2}.
\end{equation}

A deep analysis on the spectrum near the catenoid throat can be accomplished as the following.
Let us look at the power series of $\frac{\epsilon^{2}}{(1-r^{2})^{2}}$ about the point $r=0$.
Up to first-order in $r$, we have
\begin{equation}\label{susy23}
  \frac{\epsilon^{2}}{(1-r^{2})^{2}}\approx \epsilon^{2}.
\end{equation}
Using the first term of  (\ref{susy23}) in (\ref{susy22}), it becomes
\begin{equation}\label{susy24}
 -(1-r^{2})  \bar{\bar{\chi}}_{1}''(r)+2r\bar{\bar{\chi}}'_{1}(r)+\left(\frac{r^{2}-2}{4(1-r^{2})}+3m\frac{r}{(1-r^{2})}+
  \frac{m^{2}+2}{(1-r^{2})}-4-\epsilon^{2}\right)\bar{\bar{\chi}}_{1}(r)=0.
\end{equation}
If we consider the functions of Sturm-Lioville type equation given in  (\ref{susy22}) and (\ref{susy24}), 
\begin{equation}\label{susy24x}
 -(1-r^{2})  \bar{\bar{\chi}}_{1}''(r)+2r\bar{\bar{\chi}}'_{1}(r)+V_i(r)\bar{\bar{\chi}}_{1} =0,~~~~i=1,2
\end{equation}
where
\begin{eqnarray}
V_1(r)&=&\frac{r^{2}-2}{4(1-r^{2})}+3m\frac{r}{(1-r^{2})}+  \label{susy23.1}
  \frac{m^{2}+2}{(1-r^{2})}-4-\frac{\epsilon^{2}}{(1-r^{2})^{2}}\\ 
  V_2(r)&=&  \frac{r^{2}-2}{4(1-r^{2})}+3m\frac{r}{(1-r^{2})}+ \label{susy23.2}
  \frac{m^{2}+2}{(1-r^{2})}-4- \epsilon^{2}   
\end{eqnarray}
In the low energies values, $V_1(r)$ and $V_2(r)$ have similar behaviour while they act different in the higher energy values  as is shown in the figure below.
  \begin{figure}[h!]
\centering
  \centering
  \includegraphics[scale=0.7]{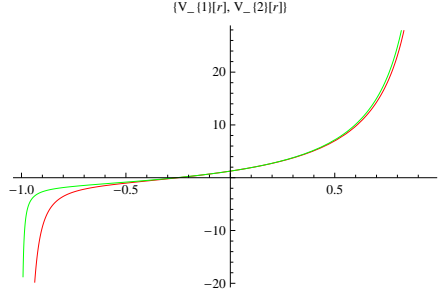}
  \caption{Graph of (\ref{susy23.1}) and (\ref{susy23.2}). $\epsilon=0.5$, $m=2$  are used for each curve. The red curve corresponds to $V_1(r)$, green one is the curve of $V_2(r)$. }.
  \label{fig:sub1}
\end{figure}%
 \begin{figure}[h!]
\centering
  \centering
  \includegraphics[scale=0.7]{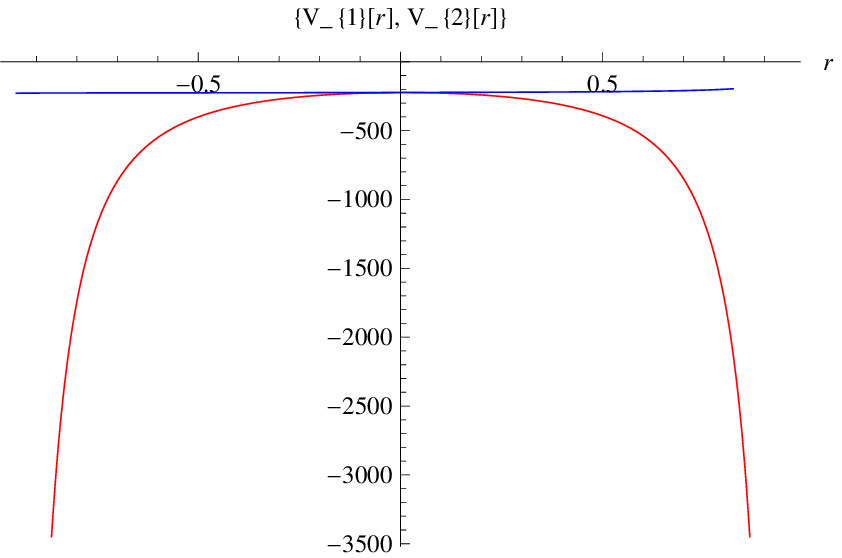}
  \caption{Graph of (\ref{susy23.1}) and (\ref{susy23.2}). $\epsilon=15$, $m=2$  are used for each curve. The red curve corresponds to $V_1(r)$, blue one is the curve of $V_2(r)$. }.
  \label{fig:sub1}
\end{figure}%
 On the other hand, (\ref{susy23.1}) and (\ref{susy23.2}) can be expanded in series when $r \rightarrow 0$ as
 \begin{eqnarray}
V_1(r)&=& -\frac{5}{2}+m^2-\epsilon^2+3mr+(\frac{7}{4}+m^2-2\epsilon^2)r^2+O(r^3)\\ 
  V_2(r)&=&  -\frac{5}{2}+m^2-\epsilon^2+3mr+(\frac{7}{4}+m^2 )r^2+O(r^3), 
\end{eqnarray}
 the system reduced into the linear potential plus harmonic oscillator system. Let us consider (\ref{susy24x}) with $V_1(r)$. A mapping $\bar{\bar{\chi}}_{1}=\frac{Z(r)}{\sqrt{1-r^2}}$ can be applied to  (\ref{susy24x}) and dividing the whole equation by $1-r^2$ leads to
\begin{equation}\label{susy2x}
-Z''(r)+U(r)Z(r)=0,
\end{equation}
where $U(r)=\frac{14-4m^2-12mr+(-31+4m^2)r^2+12mr^3+17r^4+4\epsilon^2}{4(-1+r^2)^3}$.  We can expand $U(r)$ in series when $r\rightarrow 0$, then, 
\begin{equation}\label{susy2xx}
-Z''(r)+\left(3mr+(-\frac{11}{4}+2m^2-3\epsilon^2)r^2-\frac{7}{2}+
m^2-\epsilon^2 \right)Z(r)=0.
\end{equation}
(\ref{susy2xx}) can also be written as
\begin{eqnarray}
-Z''(r)&+& U_1(r) Z(r)=\epsilon^2 Z(r),\\ \label{susy3x}
U_1&=& 3mr+\left(-\frac{11}{4}+2m^2-3\epsilon^2\right)r^2-\frac{7}{2}+ m^2 \label{susy3xx}
\end{eqnarray}
We note that (\ref{susy3xx}) is known as energy dependent potentials in the literature becasue of it's coefficient including energy parameter \cite{edp}. Moreover, it is important that $2m^2 > \frac{11}{4}+3\epsilon^2  $ which makes the coefficient of $r^2$ term as positive due to the physical solutions. The solutions of (\ref{susy3x}) can be found as
\begin{equation}\label{susy3xxx}
Z_n(r)=D_n\left(\frac{6m+8m^2r-(11+12\epsilon^2)r}{(-11+8m^2-12\epsilon^2)^{3/4}}\right)
\end{equation}
where $D_n(ar)$ are the parabolic cylinder functions and $\epsilon^2$ satisfies the relationship
\begin{equation}\label{susy4x}
\epsilon^2=\frac{1}{2f(\epsilon)^3/4}\left(-16m^4+(11+12\epsilon^2)(-7+f(\epsilon))-8m^2(-12-3\epsilon^2+f(\epsilon))\right)=n,
\end{equation}
here, $f(\epsilon)=\sqrt{-11+8m^2-12\epsilon^2}$. We are not giving the solutions of (\ref{susy4x}) which are very huge, one can compute the numerical energy eigenvalues. In the recent paper, our interest is fixing the $\epsilon^2$ term in (\ref{susy22})   without sacrificing more originality. Hence, let us make another transformation for the function $\bar{\bar{\chi}}_{1}(r)$ which leads to a transformation of (\ref{susy24}) into a Jacobi type differential equation
\begin{equation}\label{susy25}
  \bar{\bar{\chi}}_{1}(r)=   (r+1)^{b}(1-r)^{a}Y(r)
\end{equation}
then, (\ref{susy24}) turns into
\begin{eqnarray}\label{susy26}
  (1-r^{2}) Y''(r)-2\left(a-b+(1+a+b)r\right)Y'(r)+ [ \frac{1}{4}(17-4a^2-4b^2)+\epsilon^{2} &+ & \\ \frac{7-16a^2+12m+4m^2}{8(r-1)}  +\frac{-7+16b^2+12m-4m^2}{8(r+1)}-b-a(1+2b) ] Y(r)=0. \nonumber
\end{eqnarray}
(\ref{susy26}) is in the form of Jacobi differential equation which is given by
\begin{eqnarray}\label{susy27}
  (1-r^{2}) y''(r)+\left(\beta-\alpha -(\alpha+\beta+2)r\right)y'(r)+ n(n+\alpha+\beta+1)y(r)=0.  \label{susy27}
\end{eqnarray}
 Matching (\ref{susy26}) and (\ref{susy27}) gives
 \begin{eqnarray}
 a&=& \frac{1}{4} \sqrt{7+12m+4m^2},\\
  b &=& \frac{1}{4} \sqrt{7-12m+4m^2}.
 \end{eqnarray}
The parametrs $a,b$ makes the rational terms in (\ref{susy26}) equal to zero. and we also obtain,
\begin{eqnarray}
 \alpha &=&  2a \\
  \beta  &=&  2b.
 \end{eqnarray}
 $\bar{\bar{\chi}}_{1}(r)$ can be obtained as
 \begin{equation}\label{susy28}
  \bar{\bar{\chi}}_{1}(r)=   (r+1)^{b}(1-r)^{a} P_{n}^{(2a, 2b)}(r),
\end{equation}
  where $P_{n}^{(2a, 2b)}(r)$ are the Jacobi polynomials. Hence, we can give the whole solutions of our system  for $n=\nu$ (\ref{susy9})
\begin{eqnarray}\label{susy35}
 \chi^{(1)}_{n}(u)=N_{1}   \left(1-\frac{u}{\sqrt{u^{2}+R^{2}}}\right)^{\frac{1}{4} \sqrt{7+12m+4m^2}-1}\left(\frac{u}{\sqrt{u^{2}+R^{2}}}+1\right)^{\frac{1}{4} \sqrt{7-12m+4m^2}-1}P^{(2a,2b)}_{n}\left(\frac{u}{\sqrt{u^{2}+R^{2}}}\right),
\end{eqnarray}
$N_1$ is the corresponding normalization constant and the energy eigenvalues can be found as
\begin{equation}
   E^{(1)}_{n}=\pm  \frac{V_F}{2\sqrt{2}R} \left[ -27+4m^2+2(M_1+M_2)+M_1M_2+8n^2+4n(M_1+M_2) \right]^{1/2},
\end{equation}
where $M_1=\sqrt{7+4m(m-3)}$ and $M_2=\sqrt{7+4m(m+3)}$. The results agree with the solutions already obtained in the literature \cite{sukhatme}. It is important to highlight here that since the energy must be real, it imposes a constraint condition between the principal quantum number $n$ and the orbital angular momentum $m$, such that the inside of the square root should be positive for physical solutions. In fig.(\ref{Fig2}a) we can see the probability density and in fig.(\ref{Fig2}b) the probability ring for $n=1$ and $m=-2$. These values for $n$ and $m$ were chosen in order to obey the constraint imposed by the realness of the energy. From the plots we can see a region of confinement of the electron around the throat of the catenoid. Similarly, for $n=3$ and $m=-2$ we can see in fig.(\ref{Fig3}) that besides the confinement region at the throat of the catenoid, we have also two other regions of confinement, one in the upper sheet and another in the lower sheet.

We can also calculate the  $\bar{\chi}^{(2)}_{n}(u)$ which is the solution of (\ref{susy010}) using (\ref{susy8}):
\small
\begin{eqnarray}
 \nonumber\chi^{(2)}_{n} &=& N_2 \frac{1}{\sqrt{E_{n+1}}}\left(\frac{d}{u}+\frac{m}{\sqrt{R^{2}+u^{2}}}\right)\chi^{(1)}_{n+1}(u)
\end{eqnarray}
\normalsize
 Considering the partner Hamiltonian $\mathcal{H}_{2}$, let us introduce the corresponding transformed partner potential $V_{2,eff}(x)$ corresponding to (\ref{susy010}).
\begin{equation}\label{susy37}
  V_{2,eff}(x)=m^{2}\sec^{2}x+m\tan x \sec x
\end{equation}
which is sharing the same energy spectrum with $\emph{H}_{1}$.\\
%%%%%%%%%%%%%%%%%%%%%%%%%%%%%%%%%%%%%%%%%%%%%%%%%%%%%%%%%%%%%%%%%%%

\begin{figure}[h!]
\centering
\begin{subfigure}{.5\textwidth}
  \centering
  \includegraphics[scale=0.7]{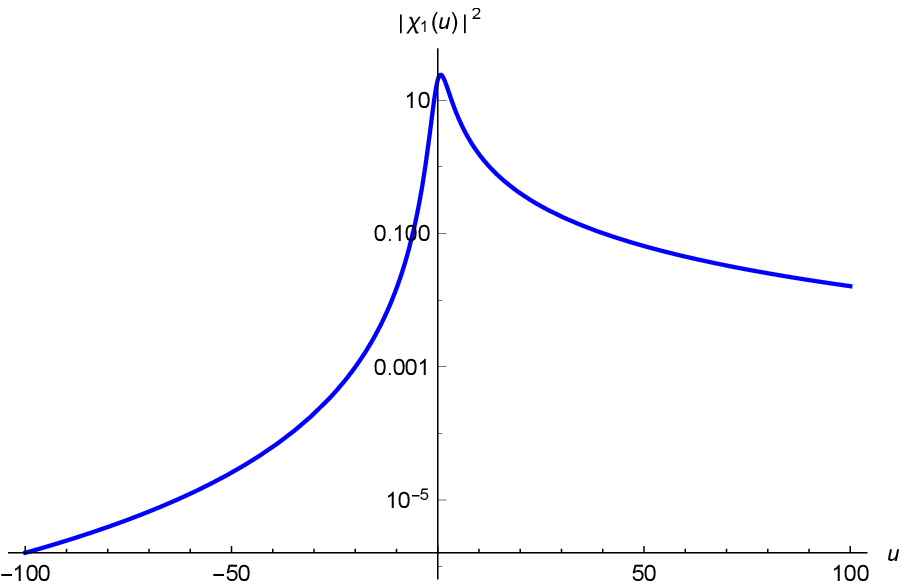}
  \caption{}
  \label{fig:sub1}
\end{subfigure}%
\begin{subfigure}{.5\textwidth}
  \centering
  \includegraphics[scale=0.7]{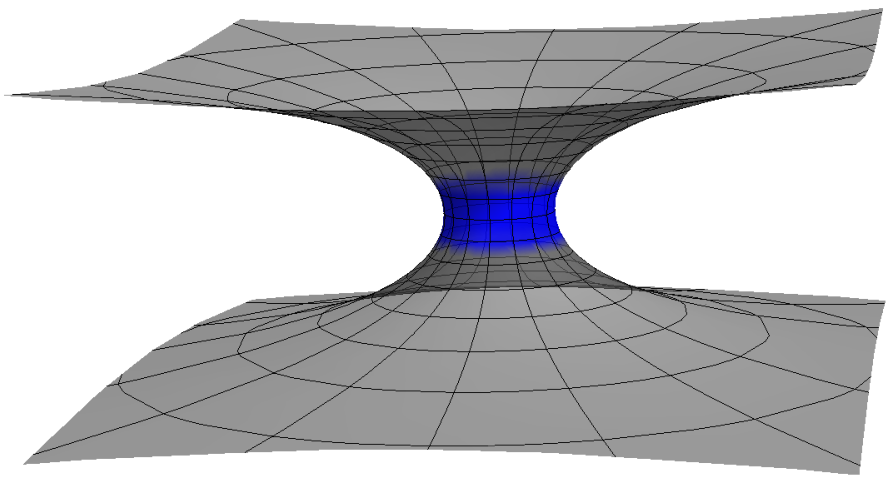}
  \caption{}
  \label{fig:sub2}
\end{subfigure}
\caption{(a) Probability density for $\chi_{n}^{(1)}(u)$ when $n=1$ and $m=-2$. (b) Probability ring for $\chi_{n}^{(1)}(u)$ when $n=1$ and $m=-2$}
\label{Fig2}
\end{figure}

\begin{figure}[ht!]
\centering
\begin{subfigure}{.5\textwidth}
  \centering
  \includegraphics[scale=0.7]{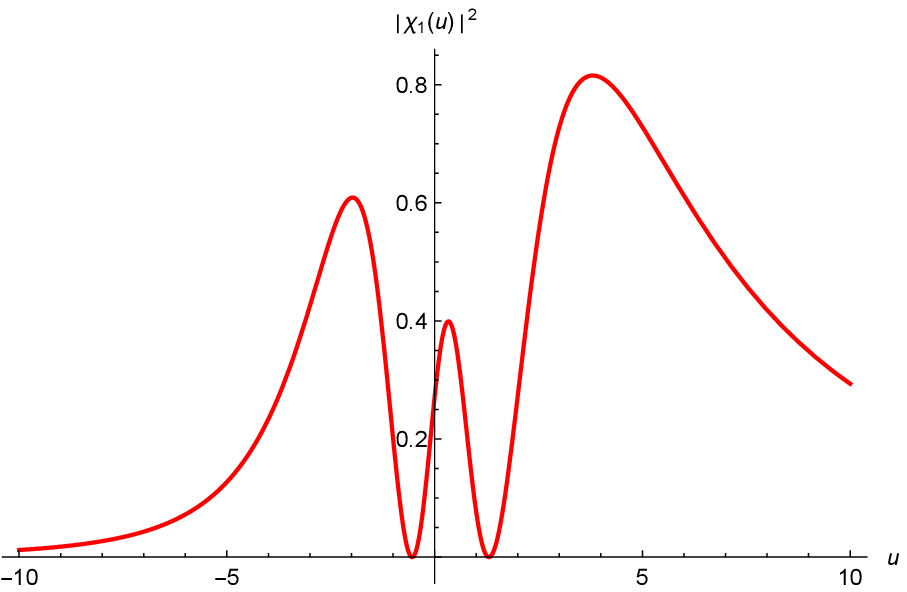}
  \caption{}
  \label{fig:sub1}
\end{subfigure}%
\begin{subfigure}{.5\textwidth}
  \centering
  \includegraphics[scale=0.7]{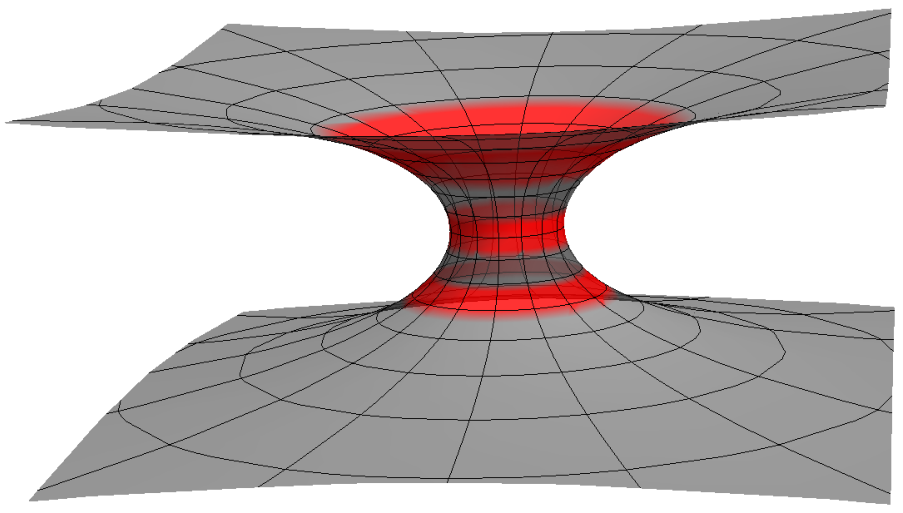}
  \caption{}
  \label{fig:sub2}
\end{subfigure}
\caption{(a) Probability density for $\chi_{n}^{(1)}(u)$ when $n=3$ and $m=-2$. (b) Probability rings for $\chi_{n}^{(1)}(u)$ when $n=3$ and $m=-2$.}
\label{Fig3}
\end{figure}

%%%%%%%%%%%%%%%%%%%%%%%%%%%%%%%%%%%%%%%%%%%%%%%%%%%%%%%%%%%%%%%%%%%%
\subsection{position-dependent Fermi velocity case }
We consider now the transformed equation for the position-dependent Fermi velocity case presented in the previous Section
\begin{equation}\label{susy38}
  -\omega''(u)+\left[\frac{m^2}{R^2+u^2}+\frac{mu}{(R^2+u^2)^{3/2}}+
  \frac{((v_{F}(u))')^{2}}{4v_{F}(u)^{2}}-\frac{1}{2v_{F}(u)}\left(v''_{F}(u)+\frac{2mv'_{F}(u)}{\sqrt{R^{2}+u^{2}}}\right)\right]
  \omega(u)=\frac{E^{2}}{v_{F}(u)^{2}}\omega(u).
\end{equation}
Using a point transformation which is $u=R \tan x$ in (\ref{susy38}) leads to
\begin{equation}\label{susy39}
  -\omega''(x)+2\tan x \omega'(x)+\Delta(x)\omega(x)=\frac{E^{2}R^{2}}{v_{F}(x)^{2}}\omega(x),
\end{equation}
where
\begin{equation}
    \Delta(x)=m^{2}\sec^{2}x+m\tan x\sec x+\frac{R \sec x}{4}\frac{v'_{F}(x)}{v_{F}(x)}-\frac{v''_{F}(x)}{2v_{F}(x)}+\frac{v'_{F}(x)}{v_{F}(x)}\tan x-\frac{mR \sec x}{v_{F}(x)}v'_{F}(x)
\end{equation}
Using $\omega(x)=\sec x \omega_1(x)$ and a suggestion on the Fermi velocity which is
\begin{equation}\label{susy390}
  v_{F}(x)=\lambda \sec^{2}x
\end{equation}
makes (\ref{susy38}) as
\begin{equation}\label{susy40}
  -\omega''_1(x)+\left[-1+(m^{2}-1)\sec^{2}x+\frac{1}{2}(2m+R-4mR)\sec x \tan x\right]\omega_1(x)=\frac{E^{2}R^{2}}{\lambda^{2}}\omega_1(x),
\end{equation}
where $\lambda$ is a real parameter. The trigonometric ansatz for the Fermi velocity could be related to periodic inhomogeneities in the graphene lattice, such as the presence of heteroatoms or topological defects. The potential function in (\ref{susy40}) is known as trigonometric Scarf-I potential in the literature whose solutions are very well-known \cite{sukhatme}. Then, the solutions of (\ref{susy40}) can be written as
\begin{equation}\label{susy41}
  \omega_1(x)=(1-\sin x)^{\frac{A-B}{2}}(1+\sin x)^{\frac{A+B}{2}}P^{(A-B-1/2,A+B-1/2)}_{n}(\sin x),
\end{equation}
where
\begin{eqnarray}\label{A}
  \nonumber A &=& \frac{1}{8(-R+m(4R-2))}\times\\
  \nonumber&&\times[-4R+8m(2R-1)+4\sqrt{2}m^{2}c-\sqrt{2}(3+c\sqrt{-3+4m^{2}(1+m-2R)(-3-2R+4m(-1+m+2R))})],  \\\\
   B &=& \frac{c}{2\sqrt{2}},\\
  c&=&\sqrt{-3+4m^{^{2}}+\sqrt{(-3+4m(1+m-2R)+2R)(-3-2R+4m(-1+m+2R))}},
\end{eqnarray}
and energy spectrum is given as
\begin{equation}\label{E}
  E^{(1)}_{n}=\pm \frac{\lambda}{R}\sqrt{(A+n)^{2}-1},~~n=0,1,2,...
\end{equation}
Also in this case, the quantity $(A+n)^{2}-1$ must be positive definite in order to assure that the energy is real. Such imposition leads to a constraint condition between the $n$ and $m$. The solutions of $\omega$ can be written in terms of $u$ as
\begin{eqnarray}\label{susy42}
  \nonumber\omega_{n}(u)=\omega^{(1)}_{n}(u)&=&N_{1}\sqrt{1+\frac{u^{2}}{R^{2}}}
  \left(1-\frac{u}{\sqrt{R^{2}+u^{2}}}\right)^{\frac{A-B}{2}}\left(1+\frac{u}{\sqrt{R^{2}+u^{2}}}\right)^{\frac{A+B}{2}}
  \times\\
  &&\times P^{(A-B-1/2,A+B-1/2)}_{n}\left(\frac{u}{\sqrt{R^{2}+u^{2}}}\right).
\end{eqnarray}
We also remind that the corresponding superpotential of the system in (\ref{susy40}) is:
\begin{equation}\label{susy43}
  W(x)=-A \tan x+B \sec x.
\end{equation}
We can introduce the partner Hamiltonian of the system given in (\ref{susy37}) as
\begin{equation}\label{susy44}
  -\omega_{2}''(u)+\left[\frac{m^2}{R^2+u^2}-\frac{mu}{(R^2+u^2)^{3/2}}-
  \frac{((v_{F}(u))')^{2}}{4v_{F}(u)^{2}}+\frac{1}{2v_{F}(u)}\left(v''_{F}(u)+\frac{2mv'_{F}(u)}{\sqrt{R^{2}+u^{2}}}\right)\right]
  \omega_{2}(u)=\frac{E^{2}}{v_{F}(u)^{2}}\omega_{2}(u).
\end{equation}
Using (\ref{susy43}) and (\ref{susy41}), we can calculate the solutions of (\ref{susy44}) which shares the same energy states (\ref{E}) with (\ref{susy37}) except the ground state:
\begin{eqnarray}\label{susy45}
\nonumber\omega_{2}(x) &\sim& \sec x \frac{(\cos\frac{x}{2}-\sin\frac{x}{2})(1-\sin x)^{A-1}}{2(\cos\frac{x}{2}+\sin\frac{x}{2})}\left\{(1+2A+n)\cos^{2}xP\left(n, \frac{1}{2}+A+B, \frac{1}{2}+A-B,\sin x\right)\right.\\
  &&\left.-2P\left(n+1,-\frac{1}{2}+A+B,-\frac{1}{2}+A-B,\sin x\right)(A-B+2A\sin x)\right\},
\end{eqnarray}
And the whole solutions read as
\begin{eqnarray}\label{susy46}
\nonumber\omega_2(u)&=&\omega^{(2)}_{n}(u)=\frac{N_2}{E^{(1)}_{n+1}}\sqrt{1+\frac{u^{2}}{R^{2}}}\frac{\sqrt{1+\frac{R}{\sqrt{R^{2}+u^{2}}}}
\left(1-\frac{u}{\sqrt{R^{2}+u^{2}}}\right)^{A-1}}{2\sqrt{1-\frac{R}{\sqrt{R^{2}+u^{2}}}}} 
  \left\{(1+2A+n) \frac{R}{\sqrt{R^{2}+u^{2}}}\right.\times\\
  \nonumber&&\times P\left(n, \frac{1}{2}+A+B, \frac{1}{2}+A-B,\frac{u}{\sqrt{R^{2}+u^{2}}}\right)- 2P\left(n+1,-\frac{1}{2}+A+B,-\frac{1}{2}+A-B,\frac{u}{\sqrt{R^{2}+u^{2}}}\right)\times\\
  &&\left.\times(A-B+2A\frac{u}{\sqrt{R^{2}+u^{2}}})\right\}
\end{eqnarray}

\section{final remarks}

We studied the Dirac equation for an electron constrained to move on a catenoid surface realized as a connection bridge between two layers of graphene. We decoupled the two components of the spinor and in order to obtain two Klein-Gordon-like equations. Analytical solutions were obtained using supersymmetric quantum mechanics for two cases, namely, the constant Fermi velocity case and the position dependent Fermi velocity case. In case of constant Fermi velocity, both zero energy modes and approximated bound state energies are obtained. The bound state solutions for the constant Fermi velocity are obtained in the limit of $r \rightarrow 0$.

For the constant Fermi velocity case, the supersymmetric quantum mechanical approach allows us to identify the potential as a very well known trigonometric Scarf-I potential. The eigenvalues of the Hamiltonian operator and the eigenfunctions were found exactly. It is important to highlight that the solutions of the system imposes a relation between the principal quantum number $n$ and the orbital angular momentum $m$ in order to guarantee the positively of the energy.

For the position-dependent Fermi velocity case, similarly, the supersymmetric quantum mechanical approach allows us to identify the potential as a very well known trigonometric Scarf-I potential. In this case exact solutions were also found under a trigonometric ansatz for the Fermi velocity. 

The paper not only presents important properties about the dynamics of an electron constrained to move on a catenoid bridge but also opens up new possibilities of investigation. The electron-phonon interaction as well as the thermodynamic properties will be addressed in a future work. 
\\
\\
\textbf{Data Availability Statement}
\\
The authors confirm that the data supporting the findings of this study are available within the article and its supplementary materials.

\section*{Acknowledgments}
\hspace{0.5cm} J.E.G.Silva thanks the Conselho Nacional de Desenvolvimento Cient\'{\i}fico e Tecnol\'{o}gico (CNPq), grants n$\textsuperscript{\underline{\scriptsize o}}$ 312356/2017-0 for financial support.

\end{document}